\documentclass[twocolumn,showpacs,preprintnumbers,amsmath,amssymb]{revtex4}
\usepackage{graphicx}
\usepackage{dcolumn}
\usepackage{bm}
\usepackage{multirow} 

\begin{document}
\title{\bf Quantum Phases of Long Range 1-D Bose-Hubbard Model: Field Theoretic and DMRG Study at Different Densities\\}
\author{\bf Manoranjan Kumar$^1$, Sujit Sarkar$^2$ and S. Ramasesha$^1$}
\address{\it 
$^1$Solid State and Structural Chemistry Unit, Indian Institute of Science, Bangalore 5600 12, India,\\
$^2$Poornaprajna Institute of Scientific Research, 4 Sadashiva Nagar, Bangalore 5600 80, India.}    
\date{\today}

\begin{abstract}
We use Abelian Bosonization and density matrix renormalization group method to study the 
effect of density on quantum phases of long range 1-D Bose-Hubbard model. We predict the existence
 of supersolid phase and also other quantum phases for this system. We have analyzed the role of long 
range interaction parameter on solitonic phase near half filling. We discuss the effect of dimerization in 
nearest neighbor hopping and interaction terms on the plateau phase at the half filling. 
\end{abstract}
\maketitle
\section{Introduction}
Different experimental and theoretical studies on superfluid
and superconducting nano-scale systems reveal a rich quantum 
phase diagram (QPD) with many interesting quantum 
phases \cite{pen,kim,rit,clark,havi1,zant,havi2,havi3}. One of 
the interesting quantum phases is the super solid (SS) phase in 
which the charge density and superconducting/superfluid phases characterized 
by diagonal and off-diagonal order coexist. The experimental 
findings and theoretical search for different quantum phases for 
cold atoms in optical lattice have revealed many interesting
 correlated phases of low dimensional bosonic systems \cite{dal,legg,pita,peth}. In 
this regard, Bose-Hubbard model with extended range interactions have 
been studied in detail to discover the different quantum phases of 
cold atoms in optical lattices \cite{dal,legg,pita,peth}. Here we study the quantum phases 
of a more general Bose-Hubbard model, namely, the Dimerized Bose-Hubbard model
(DBH) with extended range interactions.
The Hamiltonian of our model system is given by:
\begin{eqnarray}
\hat H & =&   - t_1 \sum_{i}  (1 + {(-1)}^{i} {{\delta}_t} ) 
(\hat b^{\dagger}_{i} \hat b_{i+1}  + h.c ) \nonumber\\ 
& & - t_2 \sum_{i} ( \hat b^{\dagger}_{i} \hat b_{i+2}  + h.c )  
 + \frac{U}{2} \sum_{i} \hat n_i( \hat n_i ~-~ 1) \nonumber\\ 
& & + {V_1} \sum_{i} (1 + {(-1)}^{i} {{\delta}_v} ) \hat n_{i} \hat n_{i+1}  
+ V_2 \sum_{i} \hat n_i \hat n_{i+2} \nonumber\\
& & -\mu \sum_{i} \hat n_{i}
\label{eq1}
\end{eqnarray}
\noindent $t_1$ and $t_2$ are the nearest-neighbor (NN) 
and next-nearest-neighbor (NNN) hopping terms respectively 
and $V_1$ and $V_2$ are NN and 
NNN interactions respectively. $U$ is the on site repulsion energy and $\mu$ is 
the chemical potential. ${\delta}_t $ and ${\delta}_v$
are the dimerization parameter for NN hopping and NN interaction 
respectively.
Manipulation of interaction range and the prediction of
different quantum phases in optical lattice loaded with cold atoms 
is more easily constructed than other 
correlated systems \cite{jack,jack2,man,kuk,sore,pac,roth,lewen,MPA,white,scal,white1,rahul}. 
Different combinations
of laser beams with inhomogeneous intensity profile and their suitable
manipulation can generate long range interactions and anisotropic 
interactions extending to a desired range. So our theoretical model (DBH)
is realizable because of the advances in the quantum state engineering of cold atoms
in optical lattices. 
We believe that our theoretical prediction may help to understand and
motivate experimentalist to design many interesting new systems. 
\section{ Model Hamiltonian and Continuum Field Theoretical Study}
Before presenting our numerical results, we briefly discuss a field theory
for the low energy and long wave length physics of DBH. 
We recast our basic Hamiltonian (Eq.1) in the spin 
language \cite{lar} to obtain;
$H_{J_1}= -2~{J_1} \sum_{i} (1 + {(-1)^{i}} {{\delta}_1}) ( S^+_{i} S^{-}_{i+1} + h.c)$,
$H_{J_2}=-2~{J_2} \sum_{i}(S^+_{i}S^{-}_{i+2} + h.c)$, 
$H_{E_{C0}}= {E_{C0}} \sum_{i} 
2 {S^Z}_{i} $,
$H_{E_{C1}}=4E_{Z_1} \sum_{i} (1 + {(-1)^{i}} {{\delta}_2})  S^Z_{i}S^{Z}_{i+1}$,
$H_{E_{C2}}=4E_{Z_2} \sum_{i} S^Z_{i}S^Z_{i+2}$.
\begin{equation}
H= H_{J_1}+H_{J_2}+H_{E_{C0}}+H_{E_{C1}}+H_{E_{C2}}
\end{equation}
The correspondence between the parameters of Eq. (1) and (2) is as follows: 
$ J_1 \sim  \langle n \rangle t_1 $ , $ J_2 \sim  \langle n \rangle t_2 $ , 
$E_{Co} \sim(U<n>~+~\mu)$  , $E_{z1} \sim V_1$, $ E_{z2} \sim V_2$  
\cite{fazi}.
One can transform the  spin chain model to a spinless fermion model through 
Jordan-Wigner transformation with the relation between the spin operators and 
the spinless fermion creation and annihilation operators given by 
$ S_n^z  =  \psi_n^{\dagger} \psi_n - 1/2 ~$,
$ S_n^-  =   \psi_n ~\exp [i \pi \sum_{j=-\infty}^{n-1} n_j]~$,
$ S_n^+  =  \psi_n^{\dagger} ~\exp [-i \pi \sum_{j=-\infty}^{n-1} n_j]~$, \cite{gia1}, 
where $n_j = \psi_j^{\dagger} \psi_j$ is the fermion number at site $j$.
We recast the spinless fermions operators in terms of field operators by the relation
\begin{eqnarray}
{\psi}(x)~=~~[e^{i k_F x} ~ {\psi}_{R}(x)~+~e^{-i k_F x} ~ {\psi}_{L}(x)]
\end{eqnarray}
where ${\psi}_{R} (x)$ and ${\psi}_{L}(x) $ describe
the second-quantized fields of right- (R) and
left- (L) moving fermions respectively.
We express the fermionic fields in terms of bosonic field by the relation 
\begin{eqnarray}
{{\psi}_{r}} (x)~=~~\frac{U_r}{\sqrt{2 \pi \alpha}}~~
e^{-i ~(r \phi (x)~-~ \theta (x))}
\end{eqnarray}
$r$ denotes the chirality of the R or L moving fermionic fields.
The operators $U_r$ commute with the bosonic field as well as  with $U_r$ of
different species but anticommute with $U_r$ of the same species.
$\phi$ field corresponds to the quantum fluctuations (bosonic) of spin and $\theta$ is the dual
field of $\phi$; $ {\phi}_{R}~=~~ \theta ~-~ \phi$ and  
$ {\phi}_{L}~=~~ \theta ~+~ \phi$.

Using the standard machinery of continuum field theory \cite{gia1},
we finally obtain the bosonized Hamiltonians
\begin{eqnarray}
H_1 & = & H_0 ~-~ \frac{{{\delta}_t}{J_1}}{2 \pi \alpha} 
\int ~\cos(2 \sqrt{K} {\phi}(x) )~dx \nonumber\\
& & - \frac{4 ( E_{Z1} - E_{Z2} )}{{(2 \pi \alpha)}^2} 
 \int \cos( 4 \sqrt{K} {\phi } (x)~)~ dx~\nonumber\\
& & + \frac{4 E_{Z1} {{\delta}_V} }{{(2 \pi \alpha)}^2}
\int {(-1)}^{x} \cos( 4 \sqrt{K} {\phi } (x)~)~ dx \nonumber\\
\label{eq5}
\end{eqnarray}
\begin{eqnarray}
H_{0}~&=&~v_0 \int_{o}^{L} \frac{dx}{2 \pi} \{ {\pi}^2 
: {\Pi}^2 : ~+~ :[ {\partial}_{x} \phi (x) ]^2 : \nonumber\\
&&~+~\frac{2 (\frac{E_{Z1}}{J_1} - 2 {J_2})}{{\pi}^2 } \nonumber\\
& & \int ~ dx~:[ {\partial}_{x} {{\phi}_L} (x) ]^2  :
+ :[ {\partial}_{x} {{\phi}_R} (x) ]^2 : \nonumber\\
&&~+~\frac{4 (\frac{E_{Z1}}{J_1} - 2 {J_2})}{{\pi}^2 }  \nonumber\\
& & \int ~ dx~
({\partial}_{x} {{\phi}_L} (x) ) ({\partial}_{x} {{\phi}_R} (x) )
\label{bos1}
\end{eqnarray}
$H_{0}$ is the gapless Tomonoga-Luttinger liquid part of the Hamiltonian
with $v_0~=\sin k_F$. 
The velocity, $v_0$, of low energy excitations is one of 
the Luttinger liquid (LL) parameters while $K$ is the other.
It reveals from Eq. 5 that for weak dimerization, there is no 
contribution from the interaction part of $H_1$, given by the last term in Eq. 5. The effective Hamiltonian obtained in this limit is 
the Hamiltonian for the saw tooth spin chain \cite{sawtooth} 
with dimerization. For strong dimerization, the Hamiltonian in 
Eq. \ref{eq5} reduces to 
\begin{eqnarray}
{H_1} ~=~\frac{4  E_{Z1}{{\delta}_V} }{{(2 \pi \alpha)}^2}
\int \sin ( 4 \sqrt{K} {\phi } (x)~)~ dx~.
\end{eqnarray}
 the second term in Eq. 5 and Eq. 7 yield a gap in the elementary excitations
of system which led to plateaus in the $\mu$ vs $\rho$ (boson density) 
in the system. In Density Matrix renormalization group (DMRG) study 
we will see evidences of plateau phases for different boson fillings and the effect 
of ${\delta}_t$ and ${\delta}_V$ on these plateaus. We will also see occurrence of 
gapped phase for several commensurate fillings in our DMRG study, in the next section.  
Here we build up a general field theoretical study to explain
the appearance of gap structure at different commensurate fillings: 
suppose we consider
a periodic potential $V(x)$ of periodicity of $'a'$
coupled to the density ${\rho} (x)$ leading to an additional term  in the 
Hamiltonian,
\begin{eqnarray}
{H_2} = \int dx V(x) {\rho} (x)
\end{eqnarray}
where $ V(x)= \sum_{r} {V_{r}} \cos(\frac{2 \pi  r x}{a})$, $r$,  an integer 
and ${\rho}(x) = [{\rho}_0 - \frac{\nabla{\phi (x)}}{\pi}] \sum_{p}
e^{2 p i (\pi {{\rho}_x} x - \phi (x)}$. Following Ref. 30 and 31, 
the non oscillatory contribution of $H_2$ arises from the 
commensurability condition 
$nd= p a$, $d$ is the mean distance between the particle, related 
to the density of the lattice. Under this condition, Hamiltonian 
for a particular value of $n$ is given by 
\begin{eqnarray}
{H_2} = {V_n} \int dx \cos (2 p \phi (x) )
\end{eqnarray}
$p =1$ is the most relevant commensurability and corresponds to one boson per site.
$p =2 $ is the next relevant commensurability, with one boson every two sites. 
For these commensurabilities sine-Gordon coupling term becomes relevant and 
system becomes gapped.  
\section{DMRG STUDY}
We now present numerical results obtained by using DMRG. 
We also compare them with the existing analytical and 
numerical results. 
\subsection{Numerical Details}
We use the Density Matrix renormalization group (DMRG) method to numerically study the QPD 
 of the Hamiltonian in Eq. 1. We employ the infinite DMRG algorithm keeping 128 dominant 
density matrix eigenvectors (DMEV) for determining $\mu$ while for the
calculation of correlation functions we use finite DMRG algorithm 
keeping the same cut-off in the number of DMEVs. Fock space of the site-boson 
is truncated to four states which allows 0, 1, 2 or 3 bosons per site. 
The length of the chain studied is 128 sites, except near 
phase boundaries, where we have used 256 sites for calculating $\mu$ and 
correlation functions. Accuracy of the method is checked by comparing the 
ground state (gs) energies, various correlation functions and charge gap from 
DMRG studies with exact diagonalization studies of small systems with upto 12 sites. We have 
also reproduced the results of earlier DMRG calculations satisfactorily 
\cite{white}. The discarded density in the DMRG calculations is less than 
$10^{-14}$ in the charge density wave (CDW) phase at $\rho=0.5$ as well as 
in the $\rho=1.0$, Mott-insulating phase. However, the discarded density 
is slightly less than $10^{-10}$ in the superfluid (SF) phase. 
We have computed the charge gap ($\Delta$) defined as 
$\Delta=\lim_{N \rightarrow \infty}\Delta(N)$; $\Delta(N)=E_N(p)+E_N(h)-2E_N(0)$, 
where $N$ is number of sites on the chain and $p$, $h$ correspond to an extra hole or 
extra particle at density $\rho$.
$E_N(0)$ is the gs energy of zero particle or hole number at the same density.  
\section{Results and Discussion}
\begin{figure}
\begin{center}
\includegraphics[height=9.0cm,width=9.0cm,angle=-90]{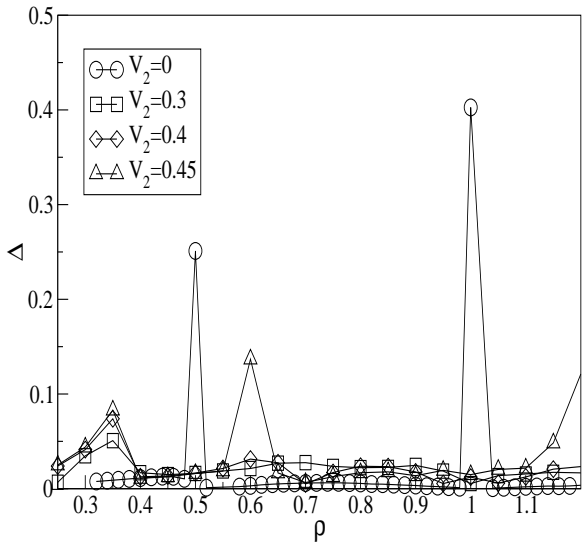}
\caption{Energy gap (${\Delta}$) vs. density ($\rho$) for different values of ${V_2}$. 
The parameters for this figure are ${t_1}=0.1$, ${t_2}=0.0$, $ U=1.0$, ${V_1}= 0.7$,
${{\delta}_t}=0$ and ${{\delta}_V}=0 $  }
\label{fig7.6}
\end{center}
\end{figure}
Fig. \ref{fig7.6} shows the variation of $\Delta$ with $\rho$ for 
different values of $V_2 $. At ${V_2} = 0$, we observe two peaks 
in the gap at the two densities, ${\rho}=0.5$ and $1$ as reported by Batrouni 
$et. al$ \cite{scal}. These peaks shift to $\approx$ 0.35 and $\approx$ 
0.65 on introducing nonzero $V_2$. Position of peaks remains the 
same for the other nonzero value of $V_2$ we have studied. We note from  
Fig. \ref{fig7.6} that the gap occurs only 
near the two commensurate fillings of 1/3 and 2/3 (when $V_2$ is 
included in the interaction) and disappears for fillings away from 
these values. This transition from gapped phase to gapless phase is the commensurate
to incommensurate transition; the latter is due to the mismatch between the
underlying periodic potential of the lattice and periodicity in the occupancy 
of the lattice.
\begin{figure}
\begin{center}
\includegraphics[height=9.0cm,width=9.0cm,angle=-90]{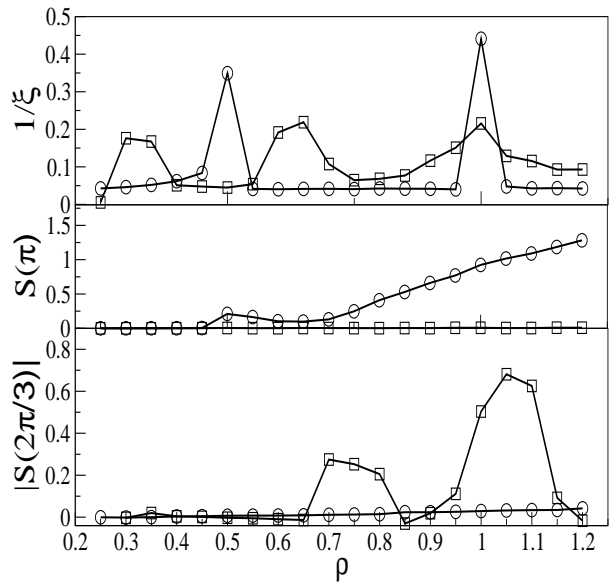}
\caption{ Dependence of inverse correlation length $1/{\xi}$, $S(\pi)$ and $|S(2\pi/3)|$ as a 
function of ${\rho}$ for $V_2=0.0$ and 0.45. Open circle corresponds to 
$V_2=0$ while square corresponds to $V_2$=0.45. Other parameter values are 
$t_1=0.1$, $U=1.0$, $V_1=0.7$, all other parameters in Eq. \ref{eq1} are 
set to zero.}
\label{fig7.7}
\end{center}
\end{figure}
Fig. \ref{fig7.7} shows the variation of inverse correlation length of the 
 density-density correlation functions $1/{\xi}$, as well as the structure 
factor $S(q)$ computed for $q=\pi$ and $2 \pi /3$ as a function of boson density, $\rho$, 
for two different $V_2$ values, namely $V_2=0$ and $V_2=0.45$. We first discuss 
the ${V_2}=0$ case. 
We note that for $V_2=0$, the inverse correlation length 
shows a peak at $\rho=0.5$ and 1.0 at which values we 
also note a gap in the system (Fig. \ref{fig7.6}). The underlying periodicity 
in the charge density at these $\rho$ values corresponds to dimerization 
as seen from large $S(\pi)$ at these fillings. We also note that for 
$V_2\neq0$, the system has vanishing $S(\pi)$ at all fillings. When $V_2$ is 
switched on, the peaks in inverse correlation length shift to $\rho=1/3$ and $\rho=2/3$; 
at these values we also observe a nonzero gap (Fig.\ref{fig7.6}) in the systems. The underlying 
charge order corresponds to a periodicity of three lattice sites for $\rho=2/3$, 
as seen from the peak in $|S(2\pi/3)|$. We also note that $|S(2\pi/3)|$ is 
vanishingly small for all $\rho$ in the case of $V_2=0$. These results are 
also in broad agreement with field theoretic results. 
\begin{figure}
\begin{center}
\includegraphics[height=9.0cm,width=9.0cm,angle=-90]{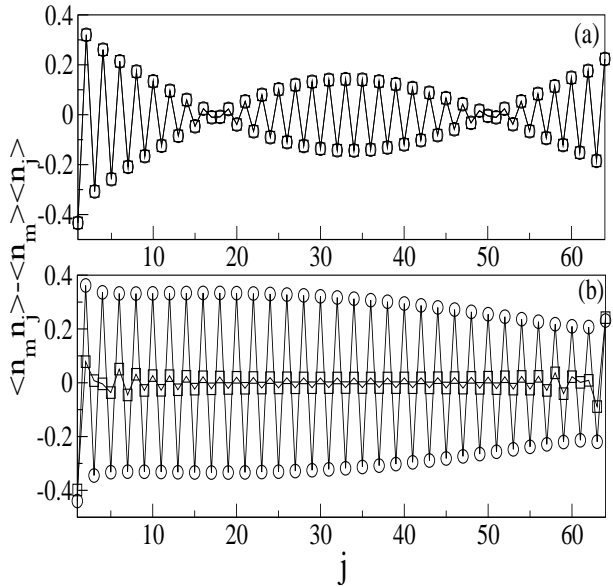}
\caption{Variation of density-density correlation with distance.
(a) one less boson than half-filling, (b) one more boson than 
half-filling. Squares and circles represent $V_2=0$ and $V_2=0.3$ respectively. }
\label{fig7.8}
\end{center}
\end{figure} 

The charge charge correlation function in the gs for hole -doping and particle 
doping are shown Figs. \ref{fig7.8}a and \ref{fig7.8}b respectively. We note that
for the case of hole doping, the 'defect' brakes up into two solitonic 
states each with charge half, for both values of $V_2$ (0.1 and 0.3) 
for $U=1$ and $V_1=0.7$. However, in case of the particle doping, we note that the 
two cases have quite different behavior. For $V_2=0.1$ we note that the 
charge-charge correlation function oscillates over the entire chain length. 
In case of $V_2=0.3$, the oscillations are damped in the middle of the chain and 
become slightly more pronounced at the ends. This behavior is akin to 
what is seen in the hole-doping case. 

\begin{figure}
\begin{center}
\includegraphics[height=7.0cm,width=7.0cm,angle=-0]{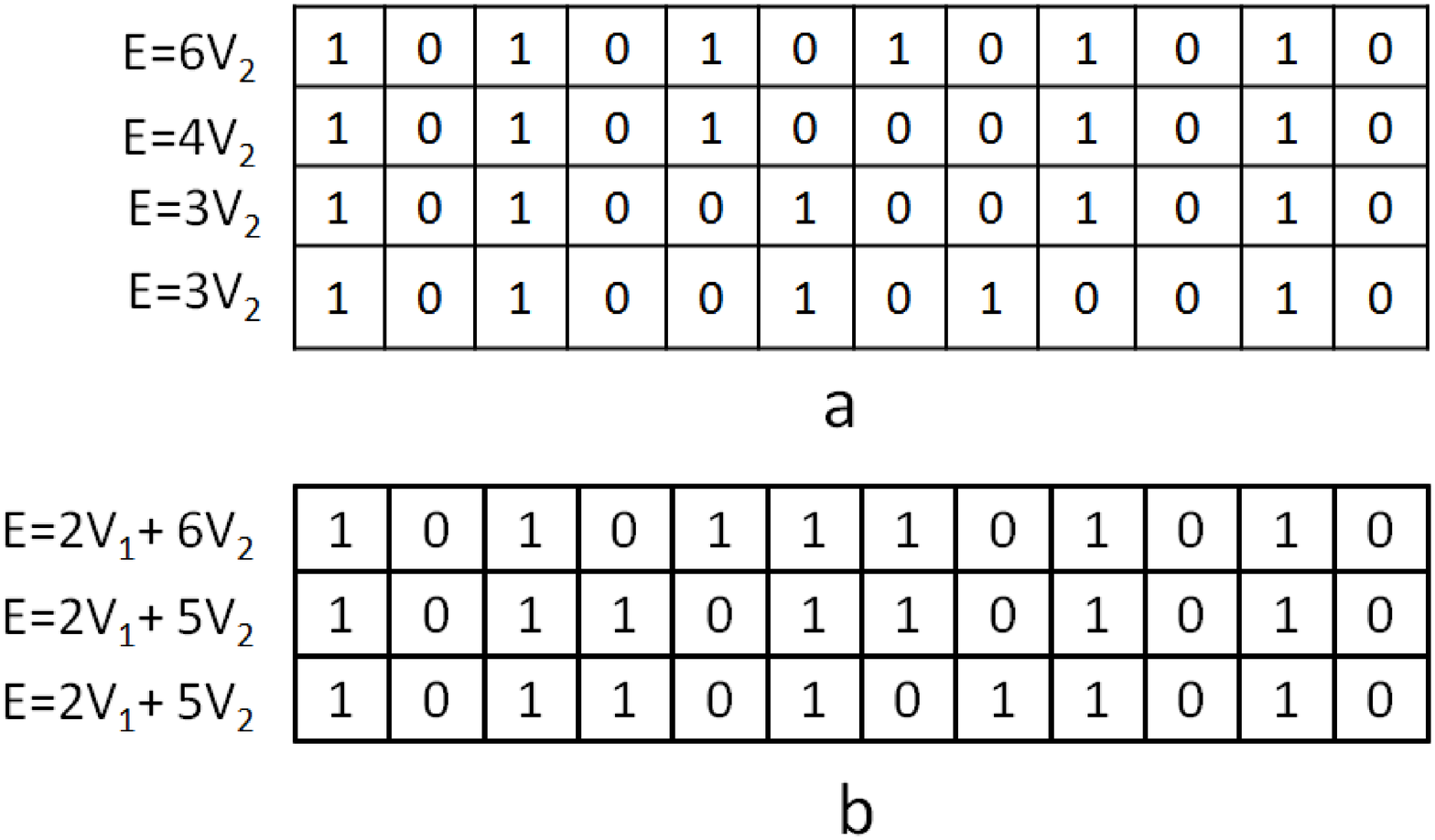}
\caption{Schematic diagrams of different configurations 
of half filled bosonic chains with one extra particle and one extra hole together 
with their energies in the $t_1=t_2=0$ limit. 
Panel (a) represents half-filling system with one extra 
hole whereas panel (b) is for one extra particle.}
\label{fig7.9}
\end{center}
\end{figure}

Physical picture for this behavior, can be arrived at from an analysis 
of the $t=0$ Hamiltonian. We find that at $\rho=0.5$ the lowest energy configuration 
is the one  in which alternate site are occupied by a single boson 
(Fig. \ref{fig7.9}a top row). On doping with a single hole we find that the energy reduces by 
$2V_2$ (Fig. \ref{fig7.9}a middle row). However, if the state with these consecutive holes 
is delocalized (Fig. \ref{fig7.9}a bottom row) then there is a further stabilization by $V_2$. Thus
the system prefers to break-up into two defects, with each defect corresponding to two 
consecutive hole sites. We can formally associate a charge half with each defect 
since two defects have been created by a single hole doping.

The case of particle-doping is slightly different. When an empty site
 is doped (Fig. \ref{fig7.9}b topline) then the energy increase is $2V_1$. 
Delocalization  of the particle, leads to a state with energy $2V_1-V_2$
( Fig. \ref{fig7.9}b middle line), which is stabilized by 
$V_2$ as in the hole-doped case. However, we can also dope a 
particle at a site which is already occupied by a boson. 
The energy increase corresponds to $U+2V_2$ in this case. Thus, we should 
observe a 3 consecutive particle state yielding a state with two 
separated consecutive particle state ( Fig. \ref{fig7.9}b middle row) 
only for $2V_1< U+2V_2$. This is exactly what we find in Fig. \ref{fig7.9}b. 
We can identify the ground state for $2V_1> U+2V_2$ as a Mott insulator 
state while that for $2V_1<U+2V_2$ corresponds to a solitonic state.
\begin{figure}
\begin{center}
\includegraphics[height=9.0cm,width=9.0cm,angle=-90]{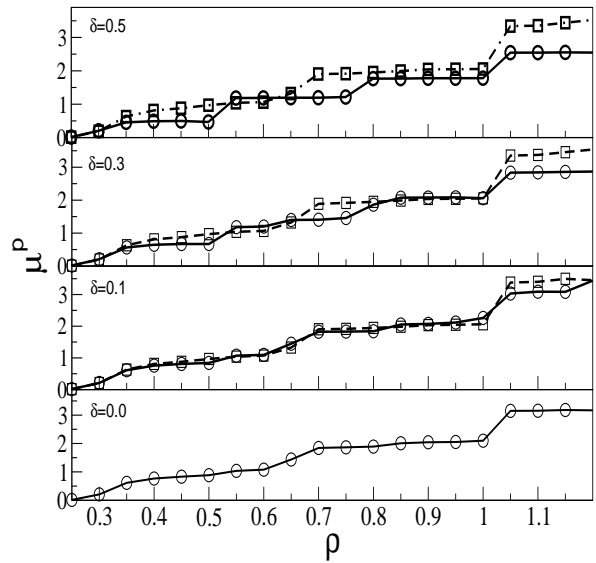}
\caption{ Density ($\rho$) vs. chemical potential ($\mu$) plot. Solid 
curve with circles are for nonzero dimerization in $t_1$ and  
dashed curves with squares are for nonzero dimerization in $V_1$. 
Parameters space for this figure are
${t_1}=0.1$, ${t_2}=0.0$, $U=1.0$, ${V_1}=0.7$, ${V_2}=0.45$. }
\label{fig7.10}
\end{center}
\end{figure}
We now turn our attention to the effect of dimerization on the phase diagrams.
We do not assume simultaneous dimerizations in both $V_1$ and $t_1$
as we wish to explore the role of each of these parameters independently.
The effect of dimerization in $t_1$ appears to smoothen the $\rho$ vs. $\mu$ 
behavior. However, the dimerization in $V_1$ seems to lead to higher jumps 
between plateaus, besides changing the value of $\rho$ at which the plateaus occur. 
It is easy to construct a real space picture for the observed $\rho$ vs $\mu$ 
behavior if we treat the transfer term as perturbation. In a model with 
next nearest neighbor interaction in 
the $t_1=0$ limit, it is possible to have a ground state with zero energy for 
all fillings, $\rho< 1/3$ for which the particle can be so distributed that the inter 
particle interaction in Eq. \ref{eq1} is zero. At, $ \rho =1/3 $, when an extra particle is added, 
there is jump in the gs energy by $(2V_2)$ which is reflected as a step in the $\rho$ vs. 
$\mu$ plot. Further addition of particle will increase the gs energy by same amount 
until $\rho< 1/2$, keeping $\mu$ constant between $\rho=1/3$ and $\rho=1/2$. This 
picture can be extended further for higher fillings. The effect of the 
transfer term is to reduce the sharpness of the jumps as well as introduce a slow 
variation in $\mu$ between jumps. For $\rho > 1/2$, the chemical potential is nearly
 $U$ for $U<2V_1$ and $2V_1$ for $U>2V_1$, when transfer term is switched on. 
Dimerization of the lattice does not significantly affect the 
$\rho$ vs. $\mu$ behavior at least up to $\rho=1$. Beyond $\rho=1$, 
the analysis is not straightforward due to the large number of occupancy 
possibilities afforded by the the bosonic system.

\noindent The physics of our system is similar to that of a sawtooth spin chain 
under a magnetic field mentioned in sec. II. Here $\rho$ of the bosonic model is replaced by 
the magnetization and $\mu$ is replaced by the magnetic field. We would 
like to give the physical explanation of the plateau state following 
the reference \cite{hone}. The energy levels of a magnetic chain can be 
labeled by the $M_s$ value of the state. When an external magnetic 
field $H$ is applied the state is stabilized by an energy $-gHM_s$. Thus if the 
$M_s$ value of the gs in the absence of an external field is zero, when the 
field is turned on, the gs switches progressively to higher values of 
$M_s$. If to begin with, the system had gaps between the lowest energy state in 
different $M_s$ sectors, then the $M_s$ value of the gs state shows jumps at
discrete values of the magnetic field. This results in plateau in the 
$M_s$ vs $H$ plot.
 
\noindent In summary, We have carried out quantum phase analysis of dimerized 
Bose-Hubbard model, emphasing quantum field theoretic treatment 
as well as DMRG method to follow the quantum phases. A real space picture of 
the system in the zero hopping limit gives clear insights into the 
nature of the quantum phases, which are also predicted by the 
quantum field in the strongly interacting limit.

{\bf Acknowledgement}: MK thanks UGC, India for financial support, SS thanks dept. 
of Physics, IISc for facilities extended. This work was supported in part by
a grant from DST (No.SR/S2/CMP-24/2003), India.      

\end{document}